\documentclass[12pt]{iopart}
\usepackage{iopams}  
\usepackage{graphicx}
\def\be{\begin{equation}}
\def\ee{\end{equation}}

\newcommand{\beq}{\begin{equation}}
\newcommand{\eeq}{\end{equation}}
\newcommand{\ber}{\begin{eqnarray}}
\newcommand{\eer}{\end{eqnarray}}
\newcommand{\berr}{\begin{eqnarray*}}
\newcommand{\eerr}{\end{eqnarray*}}

\begin{document}

\title{ 
Progress in finite temperature 
lattice QCD
}

\author{P\'eter Petreczky 
}
\address{ 
Department of Physics and RIKEN-BNL Research Center, Brookhaven
National Laboratory, Upton, New York, 11973
}

\begin{abstract}
I review recent progress in finite temperature lattice calculations, including the
determination of the transition temperature, equation of state, screening of static
quarks and meson spectral functions.
\end{abstract}




%
\section{Introduction}
\label{intro}

One expects that at sufficiently high
temperatures and densities the strongly interacting matter undergoes a
transition to a new state, where quarks and gluons are no longer confined in
hadrons, and which is therefore often referred to as a deconfined phase
or Quark Gluon Plasma (QGP).
The main goal of heavy ion experiments is to
create such form of matter and study its properties.
We would like to know at which temperature the transition takes place and what
is the nature of the transition as well the properties of the deconfined
phase, equation of state, static screening lengths, transport properties etc.
Lattice QCD can provide first principle calculation of the transition temperature,
equation of state and static screening lengths (see Ref. \cite{sewm06,lat06}) for recent reviews. 
Calculation of transport coefficients remains an open challenge for lattice QCD 
(see discussion in Refs. \cite{aarts,derek}).
 
 One of the most interesting question for the lattice
is the question about the nature of the finite temperature transition
and the value of the temperature $T_c$ where it takes place.
For very heavy quarks we have a 1st order deconfining transition.
In the
case of QCD with three degenerate flavors of quarks we expect a 1st order
chiral transition for sufficiently small quark masses.
In other cases there is no true phase transition but just a rapid
crossover.
Lattice simulations of 3 flavor QCD with improved staggered quarks (p4) using
$N_{\tau}=4$ lattices indicate that
the transition is first order only for very small quark masses,
corresponding to pseudo-scalar meson masses of about $60$ MeV
\cite{karschlat03}.
A recent study of the transition using effective models
of QCD  resulted in a similar estimate for the boundary in the quark mass
plane, where the transition is 1st order \cite{szepzs}.
This makes it unlikely that for the interesting case of one heavier strange
quark and two light $u,d$ quarks, corresponding to $140$ MeV pion, the
transition is 1st order. However, calculations with unimproved staggered
quarks suggest that the transition is 1st order for pseudo-scalar
meson mass of about $300$ MeV \cite{norman}.
Thus the effect of the improvement is
significant and we may expect that the improvement of flavor symmetry,
which is broken in the staggered formulation, is very important.
But even when using improved staggered fermions 
it is necessary to do the calculations at several
lattice spacings in order to establish the continuum limit.
Recently,  extensive calculations have been done to clarify the nature
of the transition in the 2+1 flavor QCD for physical quark masses using
$N_t=4,~6,~8$ and $10$ lattices.
These calculations were done using the
so-called $stout$ improved staggered fermion formulations which improves
the flavor symmetry of staggered fermions but not the rotational symmetry,  
The result of this study was
that the transition is not a true phase transition but only a rapid
crossover \cite{nature}. New calculations with stout action indicate that
only for quark masses about ten times smaller than the physical quark mass
the transition could be first order \cite{endrodi}.
Even-though there is no true phase transition in  
QCD thermodynamic observables change rapidly in a small
temperature interval and the value of
the transition temperature plays an important role.
The flavor and quark mass dependence of
many thermodynamic quantities is largely determined by the flavor and
quark mass dependence of $T_c$. For example, the pressure normalized by
its ideal gas value for pure gauge theory, 2 flavor, 2+1 flavor and 3 flavor
QCD shows almost universal behavior as function of $T/T_c$ \cite{cargese}.

The chiral condensate $\langle \bar \psi \psi \rangle$ and the expectation value of the Polyakov loop $\langle L \rangle$
are order parameters in the limit of vanishing and infinite quark masses respectively. However, also for finite values of
the quark masses they show a rapid change in vicinity of the transition temperature. 
Therefore they can be used to locate the transition temperature. The fluctuations of the chiral condensate and Polyakov loop
have a peak at the transition temperature. The location of this peak has been used to define the
transition temperature in the calculations with p4 action on lattices with temporal extent $N_{\tau}=4$ and $6$ for several
values of the qurk mass \cite{us06}. The combined continuum and chiral extrapolation then gives the value $T_c=192(7)(4)$MeV.
In this calculations the lattice spacing has been fixed by the Sommer parameter $r_0=0.469(7)$fm \cite{gray}. The last
error in the above value of $T_c$ corresponds to the estimated systematic error in the extrapolation.
Recently the transition temperature has been
determined using the so-called $stout$ staggered action and $N_t=4,~6,~8$ and $10$
\cite{fodorplb}. The deconfinement temperature has been found to be 
$176(3)(4)$ MeV \cite{fodorplb}. The central 
value is considerably smaller than the
one obtained with $p4$ action but taking into account the errors the deviation
is not very significant. 
The authors of Ref. \cite{fodorplb} use a different definition of the chiral susceptibility which resulted
in the chiral transition temperature of $T_{chiral}=151(3)(3)$MeV. Using the
chiral susceptibility defined above would result in a larger value of the
transition temperature.  
Although the continuum extrapolation using only $N_{\tau}=4$ and $6$ is not completely reliable new 
calculations by {\em HotQCD Collaboration} using $N_{\tau}=8$ lattices give quite similar results
for the Polyakov loop, chiral condensate and strangeness susceptibility as earlier calculations with
$N_{\tau}=4$ and $6$ lattices both for the p4 and asqtad action \cite{hot}.

\begin{figure}
\includegraphics[width=7cm]{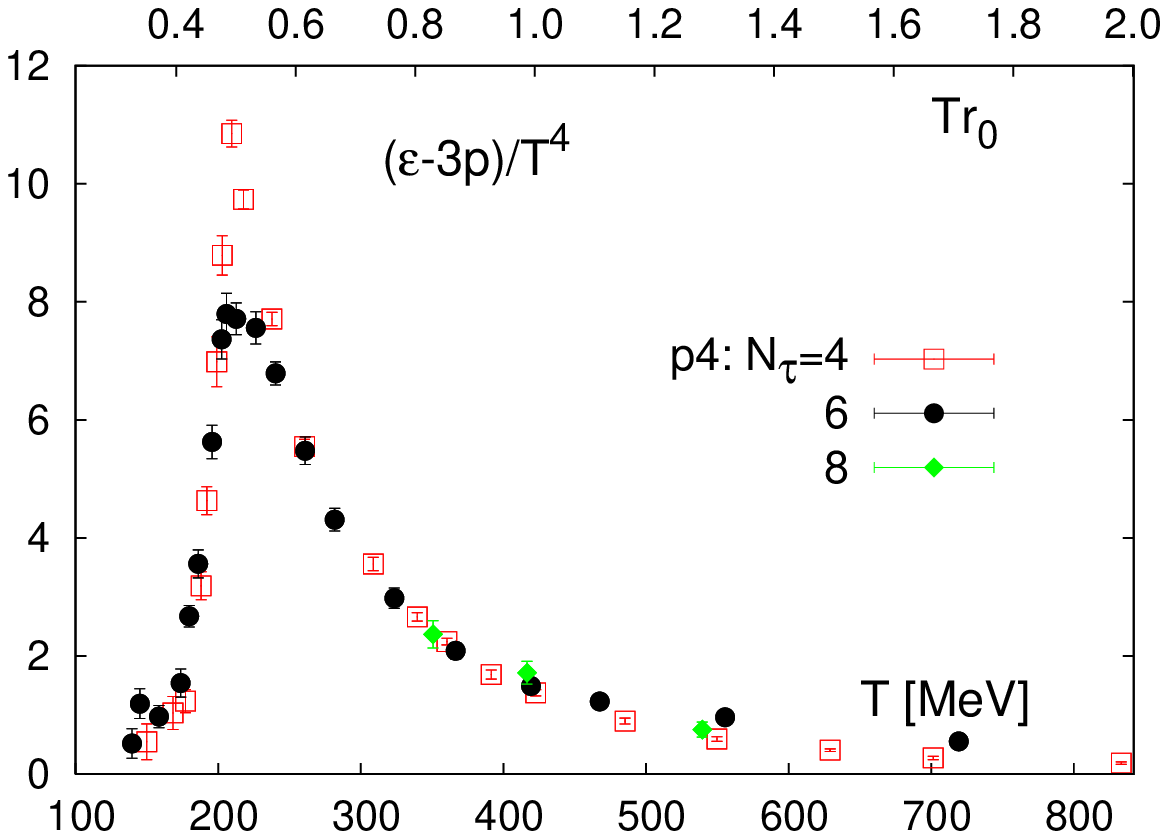}
\includegraphics[width=7cm]{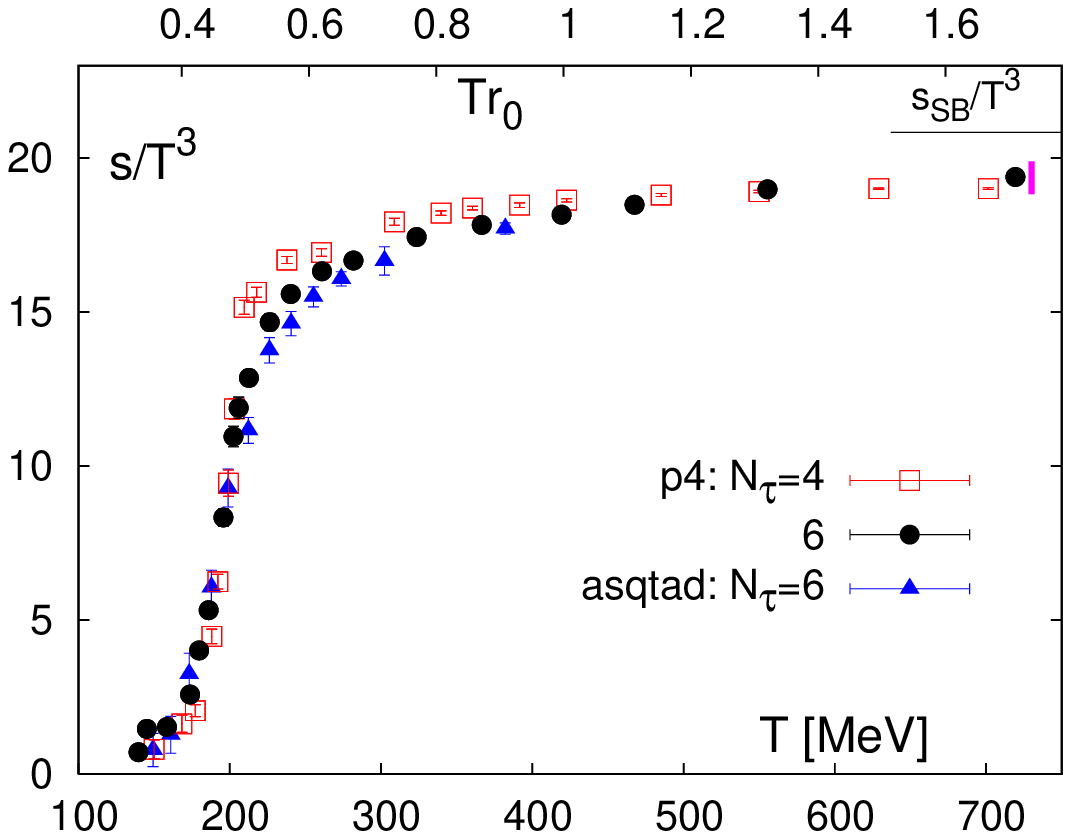}
\caption{ 
The interaction measure calculated (left) and the entropy density (right) for the p4 action \cite{eos_pap}.}
\label{fig:e-3p}
\end{figure}

Lattice calculations of equation of state were started some twenty years ago. In the case of QCD without dynamical
quarks the problem has been solved, i.e. the equation of state has been calculated in the continuum limit \cite{boyd96}.
At temperatures of about $4T_c$ the deviation from the ideal gas value is only about $15\%$ suggesting that quark gluon
plasma at this temperate is weakly interacting. Perturbative expansion of the pressure, however, showed very poor
convergence at this temperature \cite{arnold}. Only through the use of new re-summed perturbative techniques it was possible
to get agreement with the lattice data \cite{scpt97,braaten,blaizot}. To get a reliable calculation of the equation of state on the lattice,
improved actions have to be used \cite{heller99,karsch00}. Recently equation of state have been calculated using p4 and asqtad
improved staggered fermion actions \cite{milc06,eos_pap}. In lattice calculation the basic
thermodynamic quantity is the trace of the energy momentum tensor. , often refered to as the interaction measure $\epsilon-3p$.
This is because it can be expressed in terms of expectation values of gauge action and quark condensates (see discussion
in Ref. \cite{eos_pap}). All other thermodynamic quantities, pressure, energy density and entropy density $s=(\epsilon+p)$ 
can be obtained from it using integration
\begin{equation}
\frac{p(T)}{T^4} -\frac{p(T_0)}{T_0^4} = \int_{T_0}^T d T' \frac{\epsilon(T')-3 p(T')}{T'^5} 
\end{equation}
The value of $T_0$ is chosen to be sufficiently small so that it corresponds to vanishing pressure
to a fairly good approximation.
In Fig. \ref{fig:e-3p} I show the interaction measure form the new calculations with p4 action on $N_{\tau}=4$ and $6$ 
lattices \cite{eos_pap}. At highest temperatures calculations with $N_{\tau}=8$ lattices have also been performed. 
As expected because of use of the improved action the difference between the $N_{\tau}=4$ and $N_{\tau}=6$ results is small.
In this figure I also show the entropy density which raises rapidly in the tempearture region $180-200$ MeV. At high temperature
it is only $10\%$ below the ideal gas limit in agreement with expectations from improved perturbative calculations. The results
from calculations with the asqtad action \cite{milc06} are also shown. These calculations agree very well with the p4 results
providing further evidence that the cutoff effects are small. 

\section{Spatial correlation functions}
To get further insight into properties of the quark gluon plasma one can study  different spatial correlation functions.
One of the most prominent feature of the quark gluon plasma is the presence of chromoelectric (Debye) screening.
The easiest way to study chromoelectric screening is to calculate the singlet free energy of static quark anti-quark pair (for recent
review on this see Ref. \cite{mehard04}),  which is expressed in term of correlation function of temporal Wilson lines
\begin{equation}
\exp(-F_1(r,T)/T)={\rm Tr} \langle W(r) W^{\dagger}(0) \rangle.
\end{equation}
$L={\rm Tr}  W$ is the Polyakov loop.
This quantity is also useful to study quarkonia binding at high temperatures 
\cite{digal01,wong,alberico,rapp,mocsy06,mocsy07}.  
In purely gluonic theory the free energy grows linearly with the separation between the heavy quark and 
anti-quark in the confined phase. In presence of dynamical quarks the free energy is saturated at some finite value 
at distances of about $1$ fm due to string breaking \cite{mehard04,kostya1,okacz04}. Above the deconfinement temperature the singlet free
energy is exponentially screened, at sufficiently large distances \cite{okacz04}, i.e.
\begin{equation}
F_1(r,T)=F_{\infty}(T)-\frac{4}{3}\frac{g^2(T)}{4 \pi r} \exp(-m_D(T) r).
\end{equation}
 The inverse screening length or equivalently the Debye screening mass $m_D$ is proportional to the temperature. In leading order
of perturbation theory it is
$m_D=\sqrt{1+N_f/3} g(T) T.$
Beyond leading order it is sensitive to the non-perturbative dynamics of the static chromomagnetic fields. 
The Debye screening mass has been calculated in pure gauge theory ($N_f=0$) \cite{okacz04} and in 2 flavor QCD  ($N_f=2$) \cite{okacz05}
as well as in 2+1 flavor QCD \cite{olaf07}.
The temperature dependence of the lattice data have been fitted with the simple 
Ansatz motivated by the leading order result : $m_D(T)=A \sqrt{1+N_f/3} g(T) T$. Here $g(T)$ is the two loop running coupling constant.
This simple form can fit the data quite well if $A \simeq 1.4-1.6$ \cite{olaf07}.
Thus the temperature dependence as well as the flavor dependence of the Debye mass is given by perturbation
theory. We also see that non-perturbative effects due to static magnetic fields significantly effect the electric screening,
resulting in about $40\%$ corrections.  However, the non-perturbative correction  is the same in full QCD and pure gauge theory.

\section{Spectral functions}
\begin{figure}
\includegraphics[width=7cm]{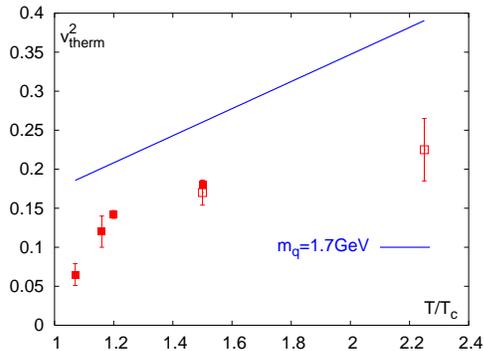}
\caption{The thermal velocity of heavy quarks estimated using anisotropic lattice data \cite{jako07} (filled symbols)
 and isotropic lattice data \cite{datta04} (open symbols).}
\label{fig:vth}
\end{figure}

Information on hadron properties at finite temperature as well as transport coefficients are encoded in different spectral functions.
In particular the fate of different quarkonium states in the quark gluon plasma can studied by calculating the corresponding quarkonium spectral functions.
On the lattice we can calculate correlation function in Euclidean time. The later is related to the spectral function via integral
relation
\begin{equation}
G(\tau, T) = \int_0^{\infty} d \omega
\sigma(\omega,T) K(\tau,\omega,T) ,~~
K(\tau,\omega,T) = \frac{\cosh(\omega(\tau-1/2
T))}{\sinh(\omega/2 T)}.
\label{eq.kernel}
\end{equation}
Given the data on the Euclidean meson correlator $G(\tau, T)$ the meson spectral function can be calculated
using the Maximum Entropy Method (MEM)  \cite{mem}. For charmonium this was done by using correlators calculated on
isotropic lattices \cite{datta02,datta04} as well as  anisotropic lattices \cite{umeda02,asakawa04,jako07} in quenched approximation.
It has been found that quarkonium correlation function in Euclidean time show only very small temperature
dependence \cite{datta04,jako07}. In other channels, namely the vector, scalar and axial-vector channel 
stronger temperature dependence was found \cite{datta04,jako07}, especially in the scalar and axial-vector channels.
The spectral functions in the pseudo-scalar and vector channels reconstructed from MEM show peak structures which may
be interpreted as a ground state peak \cite{umeda02,asakawa04,datta04}. Together with the weak temperature dependence
of the correlation functions this was taken as strong indication that the 1S charmonia ($\eta_c$ and $J/\psi$) survive
in the deconfined phase to temperatures as high as $1.6T_c$ \cite{umeda02,asakawa04,datta04}. A detailed study of
the systematic effects show, however, that the reconstruction of the charmonium spectral function is not reliable
at high temperatures \cite{jako07}, in particular the presence of peaks corresponding to bound states cannot be
reliably established. The only statement that can be is that the spectral function does not show significant changes 
withing errors of the calculations. Recently quarkonium spectral functions have been studied using potential models
and lattice data for the free energy of static quark anti-quark pair \cite{mocsy07}. These calculations show that all
charmonia states are dissolved  at temperatures smaller than $1.5T_c$, but the Euclidean correlators do not show
significant changes and are in fairly good agreement with available lattice data both for charmonium \cite{datta04,jako07}
and bottomonium \cite{jako07,dattapanic05}. This is due to the fact that even in absence of bound states quarkonium spectral functions
show significant enhancement in the threshold region \cite{mocsy07}.  Therefore previous statements about quarkonia
survival at high temperatures have to be revisited. The large enhancement of the quarkonium correlators above deconfinement in the scalar and axial-vector
channel can be understood in terms of the zero mode contribution \cite{mocsy07,umeda07} and not due to the dissolution of
the $1P$ states as previously thought. Similar, though smaller in magnitude, enhancement of quarkonium correlators due to zero mode 
is seen also in the vector channel \cite{jako07}. Here it is related to heavy quark transport \cite{derek,mocsy06}.
In the vector channel the spectral function at very small frequency, i.e. the transport contribution  is given by \cite{derek}
\begin{equation}
\sigma_{ii}^{low}(\omega)=\chi_q(T) v_{therm}^2 \frac{1}{\pi} \frac{ \eta \omega}{\omega^2+\eta^2},
\end{equation}
with $\chi_q(T)$ being the quark number susceptibility for charm or bottom quarks and $v_{therm}$ is their thermal velocity.
The width of the low energy transport contribution of the spectral functions is $\eta=T/(m_q  D)$ with $m_q$ being the
heavy quark mass and $D$ is the heavy quark diffusion constant \cite{derek}. Since $T/m_q$ is large $\eta$ is quite
small and the transport peak in the spectral functions gives an almost constant contribution to the correlator. Deviation from 
the constant gives information about the value of $D$. However, existing lattice data are not precise enough to
constrain its value. Existing lattice data can provide information about the thermal velocity of heavy quarks. On
the lattice one can also calculate the temporal component of the vector correlator which is stricly constant
$G_{00}(\tau)=-\chi_q (T) T$ \cite{derek} because of charge conservation. Therefore calculating the transport
contribtion in the vector channel and deviding it by the temporal vector correlators gives an estimate of the
thermal velocity of the heavy quarks. In Fig. \ref{fig:vth} I show the estimated thermal velocity of the charm quarks
using lattice data from isotropic and anisotropic lattices. 

Note that for free quarks the thermal velocity squared is just $T/m_q$. 
As one can see from the figure    the lattice data indicate significantly smaller thermal velocity even assuming a fairly
large value for the effective quark mass $m_c=1.7$GeV. Furthermore, we see a dramatic decrease of the thermal
velocity at temperatures close to the transition temperature.

Although the above mentioned lattice studies were performed in quenched
approximation we do not expect the picture to change when dynamical quarks are included in the calculations since
recent calculations in 2-flavor QCD show very similar temperature dependence of charmonium correlators \cite{aarts07}.

The spectral function for light mesons as well as the spectral function of the energy momentum tensor has been calculated on the lattice
in quenched approximation \cite{karsch02,asakawaqm02,aarts_el,meyer}. However, unlike in the quarkonia case 
the systematic errors in these calculations are not well understood. 

\section{Summary}

Significant progress has been achived in lattice calculations of thermodynamic quantities using improved staggered fermions.
Pressure, energy density and entropy desnity can be reliably calculated at high temperatures when improved actions are
used. Different lattice calculations show that for the physical quark masses the transition to the deconfined phase is not
a true phase transition but a crossover. There is some controversy, however, concerning the location of the crossover.
Lattice calculations provide detailed information about screening of static quarks which is important for the fate of
heavy quarkonia in the quark gluon plasma. Some progress has been made in calculating spectral functions on
the lattice, however, much more work is needed in this case. One interestimg result have been obtained 
for the thermal velocity of heavy quarks. 

\section*{Acknowledgments}
This work was supported by U.S. Department of Energy under
Contract No. DE-AC02-98CH10886.

\vfill\eject

\vskip0.5truecm

  
\vskip0.5truecm
\begin{thebibliography}{99}

\bibitem{sewm06}
  P.~Petreczky,
  Nucl.\ Phys.\  A {\bf 785}, 10 (2007)

\bibitem{lat06}
  U.~M.~Heller,
  PoS {\bf LAT2006}, 011 (2006)

\bibitem{aarts}
  G.~Aarts and J.~M.~Martinez Resco,
  JHEP {\bf 0204}, 053 (2002)

\bibitem{derek}
  P.~Petreczky and D.~Teaney,
  Phys.\ Rev.\  D {\bf 73}, 014508 (2006)

\bibitem{milc04}
  C.~Bernard {\it et al.}  [MILC Collaboration],
  Phys.\ Rev.\ D {\bf 71}, 034504 (2005)

\bibitem{fodor05}
  Y.~Aoki, Z.~Fodor, S.~D.~Katz and K.~K.~Szabo,
  JHEP {\bf 0601}, 089 (2006)

\bibitem{us06}
  M.~Cheng {\it et al.},
  Phys.\ Rev.\  D {\bf 74}, 054507 (2006)

\bibitem{szepzs}
  T.~Herpay, A.~Patk\'os, Z.~Sz\'ep and P.~Szepfalusy,
  Phys.\ Rev.\ D {\bf 71}, 125017 (2005)

 
\bibitem{cargese}
F. Karsch, Lect.  Notes Phys. {\bf 583}, 209 (2002)
                                                                                                                                      
\bibitem{karschlat03}
  F.~Karsch, et al.,
  Nucl.\ Phys.\ Proc.\ Suppl.\  {\bf 129}, 614 (2004)

\bibitem{norman}
C. Schmidt, Nucl. Phys. B (Proc. Suppl. ) {\bf 119}, 517 (2003);  N.H. Christ and X. Liao, Nucl. Phys.
 B (Proc. Suppl.) {\bf 119}, 514 (2003)
                                                                                                                                      
\bibitem{nature}
  Y.~Aoki, G.~Endrodi, Z.~Fodor, S.~D.~Katz and K.~K.~Szabo,
  Nature {\bf 443}, 675 (2006)
  [arXiv:hep-lat/0611014].

\bibitem{endrodi}
  G.~Endrodi, Z.~Fodor, S.~D.~Katz and K.~K.~Szabo,
  arXiv:0710.0998 [hep-lat].


\bibitem{okacz02}
  O.~Kaczmarek, F.~Karsch, P.~Petreczky and F.~Zantow,
  Phys.\ Lett.\  B {\bf 543}, 41 (2002)

\bibitem{kostya}
K. Petrov [ RBC-Bielefeld Collaboration], hep-lat/0610041



\bibitem{gray}
  A.~Gray et al.,
  Phys.\ Rev.\  D {\bf 72}, 094507 (2005)


\bibitem{milc01}
C. Bernard et al., Phys. Rev. D {\bf 64} (2001) 054506

\bibitem{karsch01}
 F.~Karsch, E.~Laermann and A.~Peikert,
  Nucl.\ Phys.\  B {\bf 605}, 579 (2001)

\bibitem{fodorplb}
Y. Aoki, Z. Fodor, S.D. Katz and K.K. Szab\'o, Phys. Lett. B {\bf 643}, 46 (2006)

\bibitem{hot}
  C.~DeTar and R.~Gupta  [HotQCD Collaboration],
  arXiv:0710.1655 [hep-lat].

\bibitem{boyd96}
  G.~Boyd, et al., 
  Nucl.\ Phys.\  B {\bf 469}, 419 (1996)

\bibitem{arnold}
  P.~Arnold and C.~X.~Zhai,
  Phys.\ Rev.\  D {\bf 50}, 7603 (1994)

\bibitem{scpt97}
  F.~Karsch, A.~Patk\'os and P.~Petreczky,
  Phys.\ Lett.\  B {\bf 401}, 69 (1997)

\bibitem{braaten}
  J.~O.~Andersen, E.~Braaten and M.~Strickland,
  Phys.\ Rev.\ Lett.\  {\bf 83}, 2139 (1999)

\bibitem{blaizot}
  J.~P.~Blaizot, E.~Iancu and A.~Rebhan,
  Phys.\ Rev.\ Lett.\  {\bf 83}, 2906 (1999)

\bibitem{heller99}
  U.~M.~Heller, F.~Karsch and B.~Sturm,
  Phys.\ Rev.\  D {\bf 60}, 114502 (1999)

\bibitem{karsch00}
  F.~Karsch, E.~Laermann and A.~Peikert,
  Phys.\ Lett.\  B {\bf 478}, 447 (2000)

\bibitem{milc06}
  C.~Bernard {\it et al.},
  Phys.\ Rev.\  D {\bf 75}, 094505 (2007)

\bibitem{eos_pap}
  M.~Cheng {\it et al.},
  arXiv:0710.0354 [hep-lat].




\bibitem{mehard04}
  P.~Petreczky,
  Eur.\ Phys.\ J.\  C {\bf 43}, 51 (2005)

\bibitem{digal01}
  S.~Digal, P.~Petreczky and H.~Satz,
  Phys.\ Lett.\  B {\bf 514}, 57 (2001);
  Phys.\ Rev.\  D {\bf 64}, 094015 (2001)

\bibitem{wong}
  C.~Y.~Wong,
  Phys.\ Rev.\  C {\bf 72}, 034906 (2005)

\bibitem{alberico}
  W.~M.~Alberico, A.~Beraudo, A.~De Pace and A.~Molinari,
  Phys.\ Rev.\  D {\bf 75}, 074009 (2007)

\bibitem{rapp}
  D.~Cabrera and R.~Rapp,
  arXiv:hep-ph/0611134.

\bibitem{mocsy06}
  A.~M\'ocsy and P.~Petreczky,
  Phys.\ Rev.\  D {\bf 73}, 074007 (2006)

\bibitem{mocsy07}
  A.~M\'ocsy and P.~Petreczky,
  arXiv:0705.2559 [hep-ph],
  arXiv:0706.2183 [hep-ph]

\bibitem{kostya1}
  P.~Petreczky and K.~Petrov,
  Phys.\ Rev.\  D {\bf 70}, 054503 (2004)

\bibitem{okacz04}
  O.~Kaczmarek, F.~Karsch, F.~Zantow and P.~Petreczky,
  Phys.\ Rev.\  D {\bf 70}, 074505 (2004)
  [Erratum-ibid.\  D {\bf 72}, 059903 (2005)]

\bibitem{okacz05}
  O.~Kaczmarek and F.~Zantow,
  Phys.\ Rev.\  D {\bf 71}, 114510 (2005)





\bibitem{olaf07}
  U.~M.~Heller, F.~Karsch and J.~Rank,
  Phys.\ Rev.\  D {\bf 57}, 1438 (1998)





\bibitem{mem}
  M.~Asakawa, T.~Hatsuda and Y.~Nakahara,
  Prog.\ Part.\ Nucl.\ Phys.\  {\bf 46}, 459 (2001)


\bibitem{datta02}
  S.~Datta, F.~Karsch, P.~Petreczky and I.~Wetzorke,
  Nucl.\ Phys.\ Proc.\ Suppl.\  {\bf 119}, 487 (2003)
 
\bibitem {datta04}
S.~Datta, F.~Karsch, P.~Petreczky and I.~Wetzorke,
Phys.\ Rev.\ D \textbf{69}, 094507 (2004)

\bibitem {umeda02}
T.~Umeda, K.~Nomura and H.~Matsufuru,
hep-lat/0211003
 
\bibitem {asakawa04}
M.~Asakawa and T.~Hatsuda,
Phys.\ Rev.\ Lett.\ \textbf{92}, 012001 (2004)

\bibitem{jako07}
  A.~Jakov\'ac, P.~Petreczky, K.~Petrov and A.~Velytsky,
  Phys.\ Rev.\  D {\bf 75}, 014506 (2007)

\bibitem{dattapanic05}
  S.~Datta, A.~Jakov\' ac, F.~Karsch and P.~Petreczky,
  AIP Conf.\ Proc.\  {\bf 842}, 35 (2006)

\bibitem{umeda07}
  T.~Umeda,
  Phys.\ Rev.\  D {\bf 75}, 094502 (2007)

\bibitem{aarts07}
  G.~Aarts, C.~Allton, M.~B.~Oktay, M.~Peardon and J.~I.~Skullerud,
  arXiv:0705.2198 [hep-lat].

\bibitem{karsch02}
  F.~Karsch, E.~Laermann, P.~Petreczky, S.~Stickan and I.~Wetzorke,
  Phys.\ Lett.\  B {\bf 530}, 147 (2002)

\bibitem{asakawaqm02}
  M.~Asakawa, T.~Hatsuda and Y.~Nakahara,
  Nucl.\ Phys.\  A {\bf 715}, 863 (2003)

\bibitem{aarts_el}
  G.~Aarts, C.~Allton, J.~Foley, S.~Hands and S.~Kim,
  Phys.\ Rev.\ Lett.\  {\bf 99}, 022002 (2007)

\bibitem{meyer}
  H.~B.~Meyer,
  arXiv:0704.1801 [hep-lat].

\end{thebibliography}
\end{document}